\documentclass[a4paper]{mykthesis}

\usepackage{graphicx}
\usepackage[numbers]{natbib}
\usepackage{amsmath}
\usepackage{amscd}
\usepackage{afterpage}
\usepackage[latin1]{inputenc}   

\def\bra#1{\mathinner{\langle{#1}|}}
\def\ket#1{\mathinner{|{#1}\rangle}}
\def\braket#1{\mathinner{\langle{#1}\rangle}}

{\catcode`\|=\active 
  \gdef\Braket#1{\begingroup \mathcode`\|32768\let|\BraVert\left<{#1}\right>\endgroup}
}
\def\BraVert{\egroup\,\mid@vertical\,\bgroup}
{\catcode`\|=\active
  \gdef\set#1{\mathinner{\lbrace\,{\mathcode`\|"8000\let|\midvert #1}\,\rbrace}}
  \gdef\Set#1{\left\{\:{\mathcode`\|"8000\let|\SetVert #1}\:\right\}}}
\def\midvert{\egroup\mid\bgroup}
\def\SetVert{\egroup\;\mid@vertical\;\bgroup}

\begingroup
 \edef\@tempa{\meaning\middle}
 \edef\@tempb{\string\middle}
\expandafter \endgroup \ifx\@tempa\@tempb
 \def\mid@vertical{\middle|}
\fi

\newcommand{\partiel}[2]{\frac{\partial #1}{\partial #2}}

\newcommand{\dd}[2]{\frac{\mathrm{d} #1}{\mathrm{d} #2}}

\title[A class file for KTH theses]{Nonlinear evolution as a possible explanation for quantum mechanical statevector reduction}
\author{Henrik Brusheim-Johansson}
\date{\today}
\type{M.Sc. Thesis}
\department{Departement of Physics}
\imprint{Lule\aa~2005}

\begin{document}

\begin{abstract}
A non-local toy-model is proposed for the purpose of modelling the ``wave function collapse'' of a two-state quantum system. The collapse is driven by a nonlinear evolution equation with an extreme sensitivity to absolute phase. It is hypothesized that the phase, or a part of it, is displaying chaotic behaviour. This chaotic behaviour can then be responsible for the indeterminacy we are experiencing for a single quantum system. Through this randomness, we no longer need the statistical ``ensemble'' behaviour to describe a single quantum system. A brief introduction to the ``measurement problem'' is also given.  
\end{abstract}

\begin{acknowledgments}
I would like thank my supervisor, Dr. Johan Hansson, for letting me do my thesis work at the Physics Departement at {Lule\aa} University, and always taking the time to answer my odd questions. \\
\begin{flushright}
  \emph{Henrik Brusheim-Johansson\\
  {Lule\aa}, {\today}}
\end{flushright}   
\end{acknowledgments}

\tableofcontents


\mainmatter

\chapter{Introduction}
\section{General quantum mechanical framework}
Quantum mechanics emerged as an answer to the need of describing the physics of atomic, and subatomic particles. In the early years, quantum mechanics (QM) was used to describe directly observable phenomena like emission and absorption spectra. However, it was assumed that the theory would evolve in analogy with classical mechanics, to describe trajectories of individual particles, i.e. electron orbits around a nucleus. The goal of describing individual trajectories were abandoned due to seemingly insurmountable difficulties. Thus Heisenberg developed his ``Matrix mechanics'' (1925), where all entities were divided into two main groups: observables and non-observables. The observables were the set of all experimentally measurable entities. Non-observables were explicitly excluded from the theory. This is a fundamental part of the ``orthodox'' Copenhagen Interpretation of QM.
The first steps towards a correct quantum theory were made by, among others de Broglie (1923). He suggested that Einstein's law for the energy of the light-quanta (photons)

\begin{equation}
E=\hbar\omega
\end{equation}

should be generalized to describe all particles. A consequence of this proposition would be the observation of diffraction and interference for all kinds of particles, not just photons. Numerous experiments have since then verified the theory.
Soon Erwin Schrödinger presented his view of QM; wave mechanics, where the description of the system was presented as a wave function. Schrödinger argued that electrons and photons were de facto waves. The wave description does agree with interference experiments, but has great difficulties in explaining the ability to count individual electrons on a photographic plate.\\
Schrödinger set out to formulate an equation describing the dynamics of a quantum mechanical system. He picked up on the wave-like behaviour of matter postulated by deBroglie and finally in 1926 published his wave equation, the linear Schrödinger equation (LSE)

\begin{equation}
i\hbar \partiel{\Psi}{t}=\mathcal{H}\Psi.
\label{Schr1}
\end{equation}  

Schrödinger strongly believed that his equation, in analogy with all other known physical theories, was describing the dynamics of a system evolving in a deterministic way at all times. The wave function was to be interpreted as a classical distribution.  
A direct consequence of the \emph{linearity} of Schrödinger's equation (\ref{Schr1}) is that the solution can be written as a superposition of individual (eigen)states corresponding to some observable quantity. For example, if $\Psi_1$ and $\Psi_2$ are two individual solutions, then  $\alpha_1\Psi_1 + \alpha_2\Psi_2$, where $\alpha_{1,2}$ are complex constants, is also an equally valid solution. Herein lies the very core of the ``measurement problem,'' since we do not observe superpositions when we carry out a measurement, rather we always see only \emph{one} definite outcome.\\
In Quantum Mechanics, we write the time dependent state vector $\ket{\Psi(t)}$ as an expansion
\begin{equation}
\ket{\Psi(t)}=\sum_n{c_n(t) \ket{\phi_n(t)}}
\label{expansion}
\end{equation}
, where the complete set $\{\ket{\phi_n(t)}\}$, constitutes an orthonormal (eigen)base, i.e. states that have individually observable eigenvalues. The $c_n$:s are complex expansion coefficients and their absolute values squared $|c_n|^2$, are the probabilities for observing the system in state $\ket{\phi_n(t)}$. The different $c_n$:s are obtained by projecting the state vector onto the corresponding eigenstates. Thus we can calculate the probabilities $|c_n|^2$ as (\ref{expansion})
\begin{equation}
 |c_n(t)|^2=|\braket{\phi_n|\Psi}|^2.
\label{Born1}
\end{equation}
This is the Born rule for calculating probabilities.
 That is, the n:th outcome will appear with a fraction $|c_n(t)|^2$ if successive measurements are made on a identically prepared system . So in this ``orthodox'' perspective, we interpret quantum mechanics as an inherently and fundamentally \emph{statistical} theory. Now for QM to be considered complete, it also should describe individual systems (one of Einsteins main criticisms of QM \cite{EPR1}). It would therefore be desirable to find a link between the results for an ensemble of systems and for those of an individual system.\\
Returning to the expansion of the state vector (\ref{expansion}). After a measurement has been carried out at $t=0$, all coefficients ($c_n$) but one ($c_k$) will have been driven to zero and the remaining one will be driven to unity such that
\begin{equation}
\ket{\Psi(0_{-})}=\sum_n{c_n(0_{-}) \ket{\phi_n(0_{-})}}\longrightarrow \ket{\phi_k(0_{+})}.
\label{measurement}
\end{equation}
We call this ``magical'' process the reduction of the state vector. It is \emph{not} described by the linear Schrödinger equation.
In other words; while we may interpret the left hand side of (\ref{measurement}) as a statistical description, the outcome of each individual measurement is unambiguous and definite. Note here that the coefficients in (\ref{expansion}) are time dependent \emph{only} for dynamical reduction theories as the one presented here.\\
 In the standard interpretation, the states continue to evolve while the coefficients are stationary after a measurement. Note that for a measurement of first kind (i.e. a subsequent measurement without evolution in between), the result will always be the same as for the preceding measurement (i.e. with unit probability).\\
 The standard view is that the system in question evolves in time in a unitary (i.e. norm preserving and reversible) way when it is not observed. But at (or about) the moment of measurement, the system will evolve in a non-linear and non-unitary way not described by the LSE! To acquire a solid physical understanding of the evolution of a quantum mechanical system one thus have to understand what a measurement really is. This is, in essence, the quantum ``measurement problem''. Max Born suggested that the wave function should be interpreted as a probability density. The probability for the system to be in a particular state is given by (\ref{Born1}), where the $\phi_n$ are eigenstates of the corresponding operator which properties (eigenvalues) are measured. Born immediately came to the conclusion that we do not know what state the system is in, just the probability for the various states, much to Schrödinger's' disliking. Born's suggestion led to the view that a system only has a definite property if the system is in an eigenstate with a corresponding eigenvalue equal to that very property. If the system isn't in an eigenstate it thus doesn't have the definite property. It is then in a superposition of several states, each one classically incompatible to any of the others. The Born rule certainly gives the probability of observing one of the individual states that constitute the superposition. This is also well proven by empirical observations. With this in mind, Born felt the need to suggest a new dynamical law:
If a system has a definite property only when it is in the corresponding eigenstate, then there can be no linear evolution at, or about the moment of measurement. 
This superposed state is certainly empirically distinguishable from a single eigenstate. Thus we see that the distinction theory-interpretation is not an easy one to make.
The von Neumann version of the matter is often referred to as the standard interpretation of QM and is based on a few principles:\\
\begin{itemize}
\item All physical states are described as elements with norm 1 in a Hilbert space. The elements, or in Dirac's notation: kets, are called state vectors. Two kets differing only in (constant) phase gives the same physical predictions, so all kets with the same phase really constitutes an equivalence class. In other words the absolute phase has no meaning in the standard interpretation. Although relative phase of course does have a meaning (interference). When a state can be written as a linear combination of kets, the state is  said to be in a superposition. The superposition can, and often do, include an infinite number of kets.  
\item All physical observables are represented by a corresponding Hermitian operator which acts on the vectors in the Hilbert space mentioned above.
\item A system has some definite physical value if and only if it (the system) can be described by an eigenvector to the corresponding operator. The measurement will result in the eigenvalue associated with the eigenvector. The probabilities for the different outcomes are given by the Born rule (\ref{Born1}).
\item The evolution of the states has \emph{two} incompatible parts:
\begin{itemize}
\item Collapse evolution: If a measurement is carried out, the system will evolve in a non-unitary, non-linear way to an eigenstate of the particular operator in question. The probability of collapsing to an eigenstate is given by the projection of the state vector onto the eigenstate (i.e. the Born rule (\ref{Born1})). Since the process is non-Unitary, the collapse is irreversible. Moreover, successive measurements of the same observable will always yield the same state (eigenvalue), that is, the wave function has collapsed from the superposition of possible states to a single and unambiguous state.
\item A system will evolve in a linear-unitary way when left undisturbed by measurements. Such that (in the Schrödinger picture)
\begin{equation}
\ket{\Psi(t)}=e^{-\frac{i}{\hbar}\mathcal{H}t}\ket{\Psi(0)}.
\label{evolution} 
\end{equation}
Note that this \emph{linear} evolution will automatically preserve superposition. A simple way to visualize the evolution of the states is that quantum mechanical ``particles'' always behave like (linear) waves when unobserved, and like classical particles (with definite properties) whenever we observe them.
 \end{itemize}
\end{itemize}
Now, considering the collapse evolution, we have three ways to go \citep{Espagnat};
\begin{itemize}
\item We embrace the standard interpretation and accept the \emph{two} incompatible types of evolution. In that case we have to describe in an objective way when each evolution applies. One can argue that the processes can be divided into reversible and non reversible classes, but then again, one has to do an objective choice. Moreover, given the postulate of the Copenhagen interpretation that a measurement is an irreducible concept will put us in a ``catch-22 situation'' if our true intention is to \emph{understand} physics, not just have a set of rules that certainly works, but does not provide us with any insight of the true nature of the processes.
\item We investigate the possibility of essentially \emph{one} evolution that governs the wave function. The state vector reduction has probably to be induced in an objective way and reduction probably has to take place through a modified evolution equation.
\item A non-collapse approach is taken \cite{Everett1}. It has been proposed that there is \emph{no} collapse. Instead the wave function branches out in an array of coexisting outcomes. We are experiencing only one of these branches ( ``Many worlds'' interpretation). 
\end{itemize}
Which category the present one will fit into is a matter of a debate. We will certainly not propose some entirely new equations. However, we strongly believe that it is the \emph{interaction} of a measurement that is fundamentally different from ``anything else''.

\section{A simple example}
Let us now look at a specific example to illustrate the problem. Albeit somewhat toy-like it is commonly used when describing the problem and will be used throughout the rest of the report. However it does have some physical relevance.\\
We will consider a system $\mathcal{S}$, which under the subject of measurement will result in either spin-up ($\ket{\uparrow}$) or spin-down ($\ket{\downarrow}$) e.g. an electron in a Stern-Gerlach experiment. Moreover, there is an observer $\mathcal{O}$, which takes on the states ``measured spin-up'' ($\ket{\\+}$) if $\mathcal{S}$ was observed in an ``up-state'' and ``measured spin-down'' ($\ket{\\-}$) if $\mathcal{S}$ was observed in a ``down-state''. Trivially, if $\mathcal{S}$ was in a state $\ket{\uparrow}$, the total $\mathcal{S}+\mathcal{O}$ wave function would read $\Psi=\ket{+}\ket{\uparrow}$. Now, although the outcomes of the experiment are restricted to the two possibilities mentioned above, quantum mechanically the system can, and will assume a superposition of both states, such that $\Psi_{\mathcal{S}}=\frac{1}{\sqrt{2}}(\ket{\uparrow}+\ket{\downarrow})$. It is worth to mention that this superposition is in no way just a mathematical artifact, but a real physical state. The physical relevance of such a state can be seen from self-interference phenomena (e.g. electron double slit experiment). Given the linear unitary evolution of the state, one would end up with something like $\Psi_{\mathcal{S}+\mathcal{O}}=\frac{1}{\sqrt{2}}(\ket{\\+}\ket{\uparrow}+\ket{\\-}\ket{\downarrow})$. In other words, we are not observing the ``up'', nor the ``down-state. We are not even reading ``both'' nor ``none'' because $\mathcal{O}$ does not have a state of its own. Rather the observer have become entangled with $\mathcal{S}$, but yet we are always getting unambiguous results when measuring the system! So now the complete incompability of the two evolutions is readily apparent, but still they both are absolutely necessary in the standard interpretation. Having two fundamentally different evolutions obviously raises the very disturbing question of when to use which evolution. The standard interpretation tells us that we use the unitary evolution when we are not measuring and the non-unitary, non-linear collapse when we are measuring. Yes true, but what and when is a ``measurement''? Given the rather harsh postulate that a measurement is a primitive term, we cannot really even hope to answer these questions, simply because we are not allowed to!\\
Now, bantering aside, putting some real effort in examining and eventually acquiring some understanding of the collapse phenomena, would be a giant leap forward in understanding the \emph{total} framework of quantum mechanics and in its extension; nature on its most fundamental level.

\chapter{Dynamical state vector reduction}

\section{In General}
\begin{figure}[hbt!]
\begin{center}
\includegraphics[scale=0.5,angle=0]{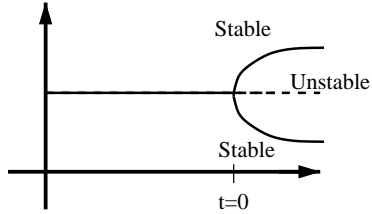} 	
\end{center}
\caption{Stability situation of collapse process. Collapse interaction at t=0, making the present (superposed) state highly unstable. Two new stable branches (possibilities) opens up due to the measurement.}		  
\label{PF}
\end{figure}

We will consider a \emph{dynamical} evolution of the state vector at all times. We emphasize that the governing equation \emph{is} the Schrödinger equation at all times. However the interaction during measurement has to be fundamentally different (non-linear) from any ``standard'' interaction. Since the model is to be viewed as a first survey of the matter, we will work in a non-relativistic Schrödinger picture (i.e. a non-local model \cite{Bell1}) . The non-locality has probably to be regarded as a fact of nature given the vast amount of empirical evidence (see Appendix A). Due to the non-locality, we will implement an action at a distance, which is mediated through the phase. The different kind of interactions (i.e. measurement) are introduced in a natural way through the nonlinear dynamics contained in the interaction Hamiltonian \cite{Hansson1}. For a measurement, the Hamiltonian scheme would be
\begin{equation}
  \mathcal{H}_{0} \Longrightarrow \mathcal{H_I}+\mathcal{H}_0 \Longrightarrow \mathcal{H}_0.
\label{Hamiltonian}
\end{equation}
Where the $\mathcal{H}_0$ denotes the Hamiltonian for the unobserved total system and $\mathcal{H_I}$ is the Hamiltonian describing the coupling due to observation. This coupling term should hence be responsible for the reduction of the state vector.\\
The nature of the interaction term has to be such that it breaks the superposition of states, i.e., it reduces the state vector such that only one $c_k$ survives the measurement (\ref{measurement}). The standard, and proven linear Schrödinger equation should be retrievable from the new model, i.e., the model should not stray too far away from the LSE. This is certainly possible if one assumes that the dynamics for the collapse process solely lies in the interaction term.
We will make the attempt of deriving a reduction theory without introducing any new variables. Instead a slightly new interpretation of the absolute phase will be made.\\
In general we never consider some direction of time in the physics laws, but just make a tacit assumption based on some previous experience. Now, in a quantum mechanical context, the time arrow has to be imposed ``by hand'', since it, as in most theories, is not evident from the governing equations. The laws are certainly time symmetric and reversible, at least for an undisturbed system, since the standard unitary evolution only corresponds to rotation in some complex vector space. Still, we experience a definite direction of time in many cases. In particular, during the collapse process, we notice this definitive direction of the time arrow, or irreversibility which certainly \emph{not} is found in the standard unitary evolution, since for \emph{any} unitary operator U, we have $UU^{\dagger}=1$. In this thesis, we will pursue the idea that the instability and hence the irreversibility is induced by the measurement interaction. Viewing the measurement as an exclusive event will likely, in accordance with orthodox quantum theory, call for an observer or a subjective reality. We just do not see any way around a subjective reality given this exclusive measurement interaction causing the irreversibility. The rather profound implication of a subjective reality is discussed in Appendix B.
Given the sensitivity induced by the measurement interaction in this model, we get a pitchfork-type of situation seen in fig.(\ref{PF}). In a way the interaction also provides new stable branches. The uncollapsed state get extremely unstable leading to a collapse within very small, but finite, timescales. 
Now, the temporal asymmetry associated with a measurement has to, logically, stem from the dynamics introduced. With an introduction of a chaotic dynamic, we will experience a dispersion of information over time, i.e., irreversibility is a consequence of a chaotic evolution in this model. If some part of the evolution equations is indeterminable (i.e. chaotic) at some stage, it could account for the indeterminism we are experiencing on an individual basis in QM. One candidate for chaotic behaviour could be the absolute phase, which through its chaotic behaviour and sensitivity is inducing and determining the reduction of the state vector.\\
In general it is said that absolute phase has no meaning in quantum mechanics \cite{Sakurai1}. However the probabilities are invariant to a phase dependence, since 
\begin{equation}
\ket{\Psi}=e^{i\theta}\ket{\Psi'} \Longrightarrow \braket{\Psi|\Psi}=\braket{\Psi'|\Psi'}.
\end{equation}
The chaos is assumed to stem from non-linear terms in the interaction coupling-Hamiltonian. Once again, making the assumption of a non-linear evolution will almost force us to look for a fundamental different ``source'' of interaction, since the linear dynamics seem to work so well for any ``standard'' quantum interaction.

\section{Derivation of our model}
Let us start with our basic wave function, i.e., the wave function in the standard model. It (and all other states) evolves in a unitary way according to
\begin{equation}
\ket{\Psi(t)}=U\ket{\Psi(0)}
\end{equation}
, where U is in general for a time-independent Hamiltonian
\begin{equation}
U=e^{\frac{-i}{\hbar}\mathcal{H}t}.
\end{equation}
For a time dependent Hamiltonian, we will have
\begin{equation}
U=e^{\frac{-i}{\hbar}\int\limits_0^\tau{\mathcal{H}(t)dt}}.
\end{equation}
We expand the wave function in some orthonormal base $\{\ket{\phi_n(t)}\}$, according to (\ref{expansion}).
To incorporate an explicit phase dependence, ${c_n}$ will be defined as
\begin{equation}
c_n=\sqrt{x_n}e^{i\theta_n}
\label{phase}
\end{equation}
, where ${\sqrt{x_n}} \in[0,1]$. Now, since we are set out to examine the \emph{collapse} behaviour, we should focus on investigating the evolution of $x_n$ in (\ref{phase}) through ${c_n}$, since under the influence of measurement, all coefficients but one should be driven to zero and the one coefficient left goes to unity.
To acquire an evolution equation for ${c_n}$, we set up a differential equation of first order with the Hamiltonians from (\ref{Hamiltonian}):
\begin{eqnarray}
\dot{c}_n&=&\dd{}{t} \braket{\phi_n(t)|\Psi(t)}\nonumber\\
&=&\braket{\dd{\phi_n(t)}{t}|\Psi(t)}+\braket{\phi_n(t)|\dd{\Psi(t)}{t}}\nonumber\\
&=&\braket{\phi_n(t)|\frac{i}{\hbar}\mathcal{H}_0+i\omega_n|\Psi(t)}+\braket{\phi_n(t)|-\frac{i}{\hbar}(\mathcal{H}_0+\mathcal{H}_I)|\Psi(t)}\nonumber\\
&=&i\omega_nc_n-\frac{i}{\hbar}\braket{\phi_n(t)|\mathcal{H}_I|\Psi(t)}\nonumber\\
&=&i\omega_nc_n-\frac{i}{\hbar}\sum_m{\braket{\phi_n(t)|\mathcal{H}_I|\phi_m(t)}\braket{\phi_m(t)|\Psi(t)}}\nonumber\\
&=&i\omega_nc_n-\frac{i}{\hbar}\sum_m{\braket{\phi_n(t)|\mathcal{H}_I|\phi_m(t)}}c_m\end{eqnarray}
,where the identity operator has been inserted in the second to last line of the equation. And $\omega_n=\braket{\phi_n(0)|\mathcal{H}_0|\phi_n(0)}$ \cite{Sakurai1,Pearle1}.
So far we have not really done anything sensational. Now let's look at the term that we are left with. What we need to ask ourselves at this point is what kind of operator it is. Linear? -not likely since it is supposed to collapse some terms and amplify one. Unitary? -hardly if we think of the collapse as an irreversible process. So assuming some nonlinear (NL), non-unitary term to the interaction, we get
\begin{equation}
\dot{c}_n=i\omega_nc_n-\frac{i}{\hbar}\sum_m{\braket{\phi_n(t)|\mathcal{H}_I^{NL}|\phi_m(t)}}c_m
\label{NLSE}
\end{equation}
To find the term $\mathcal{H}_I^{NL}$ is thus the crux of the matter.\\
Now, using (\ref{phase}), (\ref{NLSE}), we get 
\begin{eqnarray}
\dot{c}_n&=&\frac{1}{2\sqrt{x_n}}\dot{x}_ne^{i\theta_n}+\sqrt{x_m}(i\dot{\theta}_ne^{i\theta_n})\nonumber\\
&=&-i\omega_n \sqrt{x_n}e^{i\theta_n}-\frac{i}{\hbar}\sum_m{\braket{\phi_n(t)|\mathcal{H}_I^{NL}|\phi_m(t)}}\sqrt{x_m}e^{i\theta_mt}.
\label{dc}
\end{eqnarray}
Rearranging (\ref{dc}), we get the system
\begin{eqnarray}
\dot{x}_n&=&2\sum_m{\braket{\phi_n|H_I^{NL}|\phi_m}\sqrt{x_m x_n}\sin\left(\theta_m-\theta_n\right)}\label{xp}\\
\dot{\theta}_n&=&-\omega_n-2\sum_m{\braket{\phi_n|H_I^{NL}|\phi_m}\sqrt{x_m x_n}\cos\left(\theta_m-\theta_n\right)}.
\label{pp}
\end{eqnarray}
If we can find a $\mathcal{H}_I^{NL}$ such that $\dot{x}_n$ is negative for all n's but one, we will have an equation that drives all coefficients but one to zero. The one term that have a positive right hand side of (\ref{xp}), say $x_k$, will be driven to unity. This would induce a dynamical collapse process.\\
To investigate the evolution of the phase is, to understate it; difficult at the present time. Most approaches includes the postulate 
\begin{equation}
\dot{\theta}_n=0,
\end{equation}
at least during the very brief time of interaction. This is obviously a very convenient approach. A key idea of this thesis is the absolute opposite. What if the phase is \emph{anything} but well behaved? What if the phase, or at least a portion of it fluctuates so violently that it is completely indeterminable (in practice), i.e., chaotic? \\
Concentrating on the evolution of the $x_n's$, we first contemplate over the requirements we have on the $\dot{x}_n's$. As mentioned nonlinearity is needed to break the superposition. Non-Hermiticity is viewed as a strong requirement, since for a Hermitian Hamiltonian, we get
\begin{equation}
U=e^{\frac{-i}{\hbar}\mathcal{H}}\Longrightarrow U^{\dagger}=e^{\frac{i}{\hbar}\mathcal{H^{\dagger}}}\Longrightarrow UU^{\dagger}=U^{\dagger}U=1
\end{equation}
, i.e., an Unitary evolution, which hardly can describe the irreversible reduction process we are examining here. It is however very difficult to combine non-Hermiticity with a conservation of the state vector norm. Since we have ($\hbar=1$)
\begin{equation}
\frac{d}{dt}\ket{\Psi}=-i\mathcal{H}\ket{\Psi}\Longrightarrow \frac{d}{dt}\bra{\Psi}=\bra{\Psi}i\mathcal{H^{\dagger}}\Longrightarrow\frac{d}{dt}\braket{\Psi|\Psi}=i\braket{\Psi|\mathcal{H^{\dagger}}-\mathcal{H}|\Psi}. 
\end{equation}
For reasons of convergence, we require critical points at $x_n=0,1$. The critical points also have to be stable for obvious reasons. Moreover, we are trying to implement a sensitivity of phase such that phase induces and determines the collapse. With this in mind along with the desire for simplicity and a nod to Occam, we postulate the following evolution equation
\begin{equation}
\dot{x}_n=f_n(\alpha_k)\alpha_nx_n(1-x^2_n).
\label{xprim}
\end{equation}
Where $\alpha_n=\frac{cos(\theta(0)_n)}{|cos(\theta(0)_n)|}$. The function $f_n(\alpha_k)$ is the coupling between states, or ``action at a distance''
\begin{equation}
f_n(\alpha_k)=\left[1-2\sum_{k\neq n}\Theta_+\left(\alpha_k\right)\right]\cdot\alpha_n+\left[1-\sum_{k\neq n}\Theta_+\left(\alpha_k\right)\right]\cdot\Theta_+\left(\sum_{k\neq n}1-\Theta_+(-\alpha_k)\right)\cdot\left[1-\alpha_n\right].
\label{ad}
\end{equation}
The $\Theta_+'s$ are Heaviside functions such that $\Theta_+(0)=0$. We note here that a sensitivity to phase is implemented through the $\alpha's$. Moreover the coefficients are correlated by the action at a distance, which is mediated through the phase ($\alpha(\theta)$).  
Of course this corresponds to a choice of the off-diagonal matrix elements:
\begin{equation}
\mathcal{H}_{I_{nm}}^{NL}=f_n(\alpha_k)\alpha_n\frac{{x_n\left(1-x^2_n\right)}}{\sqrt{x_nx_m}\sin\left(\theta_n-\theta_m\right)}
\label{ofdiag}
\end{equation}
Clearly, having higher order terms is not beneficial because that would only slow the collapse process. The cosine fraction ($\alpha_n$) is responsible for the (extreme) sensitivity to phase. We see that the sensitivity is extreme around $\theta_n=\pm\frac{\pi}{2}$.
Note here that the equation ``register'' the phase at $t\geq0$ and then completely ``cuts away'' from the phase evolution. This corresponds to the realization of a particle from a wave i.e. a wave has a phase, a particle does not possess an intrinsic phase. So in this model, the collapse is induced, and ultimately; determined, by the (absolute) phase. The nonlinearity merely drives the evolution to the eigenstates.\\
We now turn to the evolution of the phase.
With the choice of off-diagonal elements in (\ref{ofdiag}), we see (\ref{pp}) that there are spikes in the angular speed at $\Delta\theta=k\pi,k\in N$. Now we still have the diagonal elements $\mathcal{H}_{I_{nn}}^{NL}$ to consider, which are only influencing the phase evolution.  
We postulate that $\mathcal{H}_{I_{nn}}^{NL}=\mathcal{H}_{I_{nn}}^{NL}\left(\theta_j\right)$ i.e. not dependent of ${x_n}$. And moreover, the evolution for ${x_n}$ should, according to the present theory completely detach from the phase for some $t>0$. This implies that (\ref{xprim}) is integrable and gives
\begin{equation}
x_n(t)=\frac{1}{\sqrt{1+\frac{1-x^2_n(0)}{x^2_n(0)}e^{-f(\alpha_k)\alpha_n t}}}.
\label{mod}
\end{equation}
For our two-state system it is readily seen that $x_1+x_2=1$ only asymptotically. This is just a consequence of the non-Hermiticity of the Hamiltonian. The remedy for this (only) asymptotic behaviour is not an easy one to find. The model probably needs further refinement to fit into the general framework of quantum mechanics as proposed in \cite{Grigorenko1}. It might just be impossible to achieve anything more that an asymptotic norm preservation in a theory involving transients.\\  We now pursue the possibility of the phase displaying chaotic behaviour. If we postulate that the absolute phase is in some sense indeterminable, the absolute phase \emph{could} thus account for the indeterminism we are experiencing in the measurement process. More specific, we assume that the phase has a non-chaotic part, namely the $\omega_n$ in (\ref{pp}), which is individual for each state, and a chaotic part which stems from the diagonal matrix elements in the interaction Hamiltonian. The chaotic part fluctuates wildly and indeterminable (in practice), even on a very small timescale and is \emph{common} for all states in the superposition. Nonetheless, the chaotic part has at all times a definite value. The physical point of this reasoning is that the well-behaved part is visible in e.g. interference phenomena. While the chaotic part is in some sense ``hidden'' for us, or cancels out in interference experiments due to the common value of the chaotic part.\\
Given this inability to determine absolute phase, we simply deduce that, from our viewpoint, the phase is randomly distributed in $[0,2\pi)$. So picking the initial condition for (\ref{xprim}) is really, in effect to pick a state from an equivalence class of vectors having the same $x^0_n$ (\ref{phase}), but $\theta_n \in[0,2\pi)$. This would thus account for the ``randomness'' in the measurement process.\\
 
Now, further elaborating on the phase concept; given the postulate that the phase is responsible for the random outcome of an individual measurement, there must exist a mechanism governing the collapse to the different states so that we can retrieve the statistical behaviour of the collapse. Modifying the cosine in (\ref{xprim}) to account for the probabilities, we get
\begin{equation}
\theta_n\rightarrow\frac{\theta_n}{2}-\beta,\hspace{1cm}\beta=\pi(x_n(0)-\frac{1}{2})
\end{equation}
Given the two-state system, we can form a new function q, to get a more holistic view of the collapse model.
\begin{eqnarray}
q&=&x_1-x_2\nonumber\\
&=&{\frac {1}{\sqrt {1+\frac{1-x^2_1(0)}{x^2_1(0)}{e^{-f_1(\alpha_k)\alpha_1t}}}}}-{\frac {1}{\sqrt {1+\frac{1-x^2_2(0)}{x^2_2(0)}{e^{-f_2(\alpha_k)\alpha_2t}}}}}
\label{qmod}\end{eqnarray}
\begin{figure}[hbt!]
\begin{center}
\includegraphics[scale=0.25,angle=270]{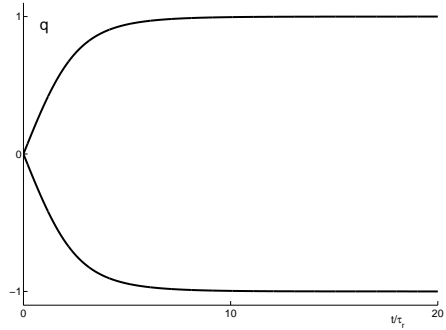} 	
\end{center}
\caption{Collapse of wave function, $\tau_r$=reduction time. A collapse to q=1 corresponds to a reduction of the state vector to $\ket{\phi_1}$, and a collapse to q=-1 corresponds to a reduction to $\ket{\phi_2}$. The strength of the interaction is inversly proportional to the reduction time. The transient part leaves an opening for experimental detection.  }		  
\end{figure}
This is our toy-model, describing the collapse of a two-state system. The model describes the collapse behaviour by driving q to $\pm1$ , i.e. to the states $\ket{\phi_{1,2}}$. Moreover it also reproduces the probabilities given by the Born rule for the measurement.\\
If we try to get a (very) rough estimate on the magnitude of the reduction time $\tau_r$, we need some ``ballpark'' energy. Let's use the energy of an incoming photon with $\lambda=400$nm. This will translate into $\tau_r=10^{-14}$s, which in turn gives us a spatial separation in the $\mu$m regime. Certanly, this estimate can give us an rough idea, but we have to keep in mind that it can differ in several orders of magnitude.

\chapter{Suggested experimental tests of dynamical state vector reduction}
\section{Predicted deviations from Malus's law}
\begin{figure}[hbt!]
\begin{center}
\includegraphics[scale=0.4]{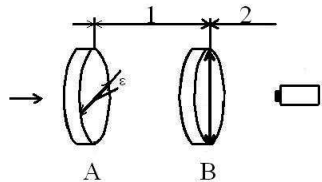}	
\end{center}
\caption{Experimental setup for testing a dynamical reduction theory against Malus's law. Double arrow indicate principal direction of polarizer} 
\label{experiment}
\end{figure}
It has been proposed to use a setup of linear polarizers to investigate the possibility of a dynamical state vector reduction \cite{Pearle2}. The wave in the domain between the first and second polarizer can be described as
\begin{equation}
\ket{\Psi}=\sin(\epsilon)\ket{\phi_1}+\cos(\epsilon)\ket{\phi_2}
\end{equation}
, where the ``1'' direction is vertical in Fig.(\ref{experiment}). Direction``2'' is perpendicular to ``1''. Now, from the experimental setup, we can see that $x(0)=\sin^2(\epsilon)$ for transmission. We calculate the expectation value for the x and compare it with the, as far as we know; precise Malus's law $(\sin^2(\epsilon))$.
\begin{eqnarray}
\braket{x}&=&\int_0^{2\pi}{x\left(t,\theta\right)d\theta}\nonumber\\
&=&\frac{\sin^2\left(\epsilon\right)}{\sqrt{1+\tan^{-2}\left(\epsilon\right)e^{-t/\tau_r}}}+\frac{\cos^2\left(\epsilon\right)}{\sqrt{1+\tan^{2}\left(\epsilon\right)e^{t/\tau_r}}}
\label{xvante}\end{eqnarray}

Plotting for some different angles $\epsilon$ and normalizing with respect to the standard Malus's law, we see some deviations, possibly inevitable for this class of reduction theories.
\begin{figure}[hbt!]
\begin{center}
\includegraphics[scale=0.3,angle=270]{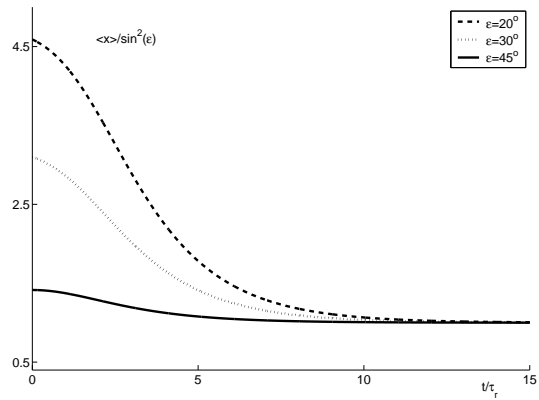}	
\end{center}
\caption{Deviation from Malus's law, $\tau_r=$reduction time. Plot is made for angles $\epsilon=20,30,45^o$, and is normalized with respect to Malus's law ($\sin^2(\epsilon)$).} 
\end{figure}
The transients seen in the figure are \emph{always} present in dynamical reduction theories with $\dot{x}_n\neq 0$. In general it is believed that one should look for inherently ``fast'' equations (\ref{xprim}) in order to minimize the transient behaviour. That is one reason to keep the lowest order term in (\ref{xprim}) ``as low as possible''. Terms of higher order are just slower. If one views standard ``Quantum Mechanics'' as an instantaneous collapse theory, it is retrievable from the present case (i.e. with the choice $\tau_r\rightarrow0$). Viewing standard QM as a ``no-collapse'' theory corresponding to a reduction time $\tau_r=\infty$, will make it non-retrievable from the present theory. The obvious way of ``avoiding'' the transient is to have a short reduction time compared to any spatial separation in Fig.(\ref{experiment}). The reduction time has thus to be at least one order of magnitude less then any timescale in which Malus's law is applicable  which would imply that the nonlinear interaction is strong. As an example; assuming that Malus's law is applicable, we consider the completely unrealistic separation of 1nm. This would hence correspond to a timescale of $10^{-19}$s and a reduction time of about $10^{-20}$s.  There have been proposals of how to determine or at least put some upper limits on the reduction time. An upper limit of $10^{-14}$s \cite{Costas1} has been placed on the reduction time $\tau_r$.

\section{Quantum phase measurement}
To put the phase concept proposed in this thesis to a test, we need a proper characterization of the quantum mechanical phase which we need to relate to an experimental situation. 
Likely, if \emph{any} detection of the chaotic part of the phase is to be made, one probably has to set up an experiment where \emph{uncorrelated} systems are used. By uncorrelated, we mean two independent sytems. Since in the proposed model, the chaotic part is common for all states it will thus cancel out in any interference experiment with correlated systems. Having said this, we \emph{do not} refute the possibility of the (chaotic) phase acting individually for all states. In that case the experiment has to be modified to study self-interference in order to try extracting some chaotic anomalies in the interference pattern. 
This type of interference is \emph{not} likely to reveal any quantative information but ``just'' a qualitative one, if at all possible.\
We try to make an extremely simple (naive) first approach to the interference of uncorrelated sources. We consider a setup with two uncorrelated sources. We assign two parameters ${y_n,\theta_n}$ to the different particles (sources). The $y_n, y'_n$ are the coordinates of the source and position on the screen. The $\theta_n$ are the individual phases.   
\begin{figure}[hbt!]
\begin{center}
\includegraphics[scale=0.3,angle=0]{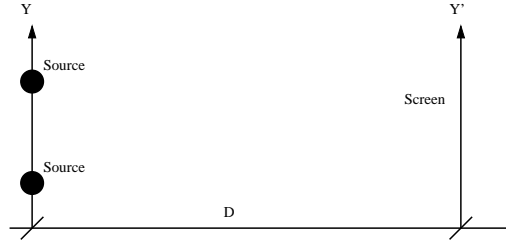}	
\end{center}
\caption{Interference scheme with variables denoted.}
\end{figure}
Now, the amplitude at the ``screen'', A, will be something like
\begin{equation}
A(y')\sim\sum_{n=1}^2e^{\imath(\theta_n+kd_n)},\hspace{1cm}d_n=\sqrt{D^2+(y_n-y')^2}.
\end{equation}
In the far-field approximation, we get
\begin{equation}
|A(y')|^2\sim|\sum_{n=1}^2e^{\imath(\theta_n+\frac{k}{2D}(y_n-y')^2)}|^2.
\label{Interfer}
\end{equation}
Although the particles are uncorrelated, we will experience an interference pattern given by (\ref{Interfer}), the idea is to look for anomalies in the interference pattern, which are non-periodic. This will obviously present some considerable obstacles. A main issue, in fact absolutely crucial as we see it, would be to distinguish noise from the proposed chaotic part of the phase i.e. a super cooled experiment seem unavoidable. Also the resolution of the experiment, or detection, has to be extreme in order to detect the predicted anomalies.\\
Almost needless to say, the naive expression (\ref{Interfer}) is hardly applicable to a real experimental situation, but only serves as an pedagogical tool. The point however, is that an interference pattern is visible for uncorrelated sources (particles). 
One possible realization would be an experiment where photons from two independent sources enters a Mach-Zehnder interferometer. Now, although this experimental setup has been used for studies of coalescence of independent photons into a two-photon state, it has been reported on interference of non-coalesced photons \cite{Bylander1}.
\begin{figure}[hbt!]
\begin{center}
\includegraphics[scale=0.3,angle=0]{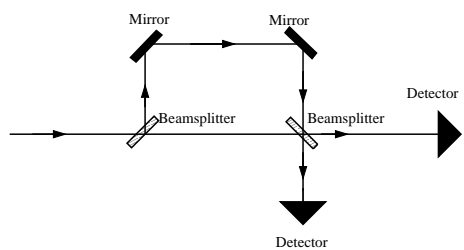}	
\end{center}
\caption{Mach-Zehnder interferometer. Independent photons from two sources coalesce in the second beam splitter to form a two photon state. A fraction of the non-coalesced photons are reported to display interference.}
\end{figure}

\chapter{Conclusion}

We have pointed out the possibility for a quantum mechanical collapse process where the total quantum evolution is governed by \emph{one} evolution equation, however, we have proposed two fundamentally different interactions. The dynamics proposed here contains nonlinear terms introduced through a non-Hermitian Hamiltonian. Reviewing von Neumann's proof (and the dismissal of the very same \cite{Bell1, Bohm1, Goldstein}) of the impossibility of hidden variable theories, suggest that a hidden variable theory should contain nonlinear relations and more important; the non-linearity breaks the superposition. Moreover the statistical behaviour is reproduced through the concept of indeterminable or chaotic absolute phase. Also, we propose that it is the phase that provide us with an ``EPR telephone'', i.e., the action at a distance, which is inherent in a non-local theory, is mediated through the phase, which in turn has to work ``above'', at least in our notion; space-time. Thus nonlinear and chaotic terms could be the remedy for the inability of quantum mechanics to describe individual behaviour of a system. Using phase as a kind of ``hidden variable'' has the advantage of not introducing any new and exotic variables into the theory. Nor have we postulated any extra evolution equations working in parallel with the usual equations $(x_n, \theta_n)$. Once the initial (Born) probabilities, $x^0_n$ are known along with the reduction time $\tau_r$, and the collapse has been induced, the theory describes the collapse process in a deterministic and causal albeit non-relativistic way. Now, choosing an appropriate $\tau_r$ is not so straightforward. Clearly, the coupling determines the strength of the nonlinear term. Or to put it in another way; the speed of the collapse process. The process cannot be ``too slow'', because that would imply the ability to observe superpositions in the ``classical'' world (non collapsed states). The possible range of $\tau_r$ is such that it is small compared to classical timescales but large compared to typical atomic and subatomic timescales. Possibly, the coupling term is \emph{not} constant. There has been suggestions of how to, at least set some boundaries on $\tau_r$ \cite{Costas1}.\\
Comparison with experimental results are made and reveals a predicted deviation from ``standard'' results (Malus's law). At sufficiently short time- or length-scales these results \emph{could} be used to dismiss reduction theories with transients. However, it is believed that there is room for refinement of the model within the given framework. An intriguing but somewhat hazy topic would be the phase evolution and the role it can play in the reduction process. Here we have repeated old sins and conveniently given the phase a position as an irreducible concept by stating the inderminability of it. A fundamental topic which has not been addressed in depth here is action at a distance. Even if this model is said to be non-relativistic one certainly has to explain the correlation between space-like separated quantum systems which are unavoidable in non-local theories like the present one.

\renewcommand\thechapter{Appendix A}
\renewcommand{\theequation}{A-\arabic{equation}}
\setcounter{equation}{0}
\renewcommand{\thesection}{A.\arabic{section}}

\appendix{}
\chapter{}
\section{On the problem with  local reality}

\subsection{Entanglement and complementary}
Entanglement is a purely quantum mechanical phenomenon. Generating, manipulating and understanding entanglement lies at the very core of quantum mechanics. It turns out that in many cases, different objects are connected to each other through space, and in a way, in time. The entangled particles thus share the same wave function although they do not interact with each other, at least not in a ``classical'' way. This is often referred to as correlation. We will see that entanglement is not consistent with a local theory where the information of a system can only be transmitted to the nearest environment, or within the relativistic light cone.
 One can suspect that entanglement is an ephemeral property. In fact experiments have shown that it is a relative robust property \cite{Alte1}. Experiments has been made where entangled photons have been fired at an opaque gold sheet. Although the photons do not penetrate the sheet, they will induce electron waves called plasmons at the surface. The plasmons travel through the sheet and re-emit a photon on the other side of the sheet. When the re-emitted photons are measured it turns out that they too are entangled! The robustness is also manifested when removing particles from a multiparticle entanglement and the remaining particles stay entangled. This is in itself a great discovery, however the robustness we observe \emph{can} have significance in a reduction context. \\
We want to stress that entanglement is not confined to just particle pairs. In fact various multi-particle entanglements has been achieved, with an eight-state entanglement recently reported \cite{8state,Catstate}.\\
As stated in the introduction to this section, entanglement and non-locality are two interconnected subjects. In fact quantum non-locality is a consequence of entanglement. We will now discuss a type of experiment which put locality and complementarity to test.

\subsection{Quantum eraser}
Let us start off considering our old friend, the double slit (Young's interference experiment). We know from experience that if individual photons pass through the slits, we will still get an interference pattern on a screen behind the slits. This is a clear manifestation of the wave-nature of the particles. One is here forced to make the conclusion that the particle passes through \emph{both} slits. If one tries to make a somewhat more direct observation by putting a detector directly at a slit, it will ``destroy'' the interference pattern and yield a more classical particle-type of detection. A twist to this experiment is that we make a delayed choice of which type of detection we will make. That is, we ``let'' the particle (photon) pass through the double slit and \emph{then}, when the photon is in flight between the slit and the screen, make a decision of the type of measurement. A local theory tells us that once the particle has passed the slit it is set to display one or the other type of behaviour. Wheeler took this \emph{gedanken experiment} to its extreme and proposed an experiment on a cosmological scale: If one views a distant galaxy as a gravitational lens, it can be considered as a double slit. We let a photon pass the galaxy and wait for a million years or so. We \emph{then} make our experiment and will thus \emph{decide} the outcome (distributions) of an event that took place millions of years ago! One possible conclusion of this experiment \emph{could} be that quantum mechanics is not confined to space-time in the (naive) way we usually think of. Various realizations of this \emph{gedanken experiments} have been made \cite{Del3,Del2,Del1}. For a quantum eraser experiment, the main idea is to make different paths distinguishable to wipe out the interference and then ``erase'' the ``which path'' information to retrieve the wave-like behaviour in accordance with quantum theory.\\
The photons are fired into a parametric down-conversion nonlinear crystal. In the crystal, some of the photons are converted into entangled photons and a vertical polarization. These photons have a lower energy than the ones incident to the crystal. Moreover, the (entangled) photons are emitted at an angle from the incident beam towards a pair of mirrors.
\begin{figure}[hbt!]
\begin{center}
\includegraphics[scale=0.4,angle=0]{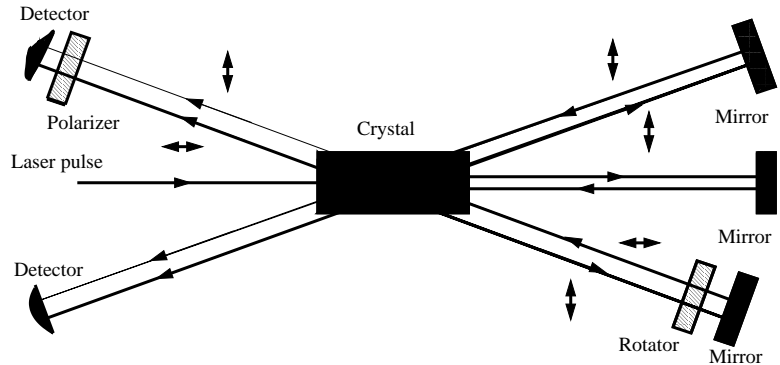} 	
\end{center}
\caption{Experimental setup of a quantum eraser experiment. The double arrows indicating polarization direction of the photons. Single arrows give the propagation direction of the photons. For more details, see for example \cite{Del4} }		  
\label{QE}
\end{figure}
Note that some photons pass through the crystal unaffected, however they will be ``re-emitted'' towards the crystal by a mirror on the opposite side of the crystal. These photons (or a fraction of them) gets down-converted on the second pass and deflects directly towards the detectors (thin lines in fig.\ref{QE}). The down-converted photons in the ``angled'' beams do not possess sufficient energy to get converted once again when reflected by the mirrors. Instead they will pass through the crystal unaffected towards the detectors. Now, we have, in effect, two double slit experiments. There is no way to identify whether the detected photon is created in the first or the second pass of the crystal. Thus we experience an interference pattern at the detectors. Now we are inserting a rotator which ``switches'' the polarization of the photon. This will cause the interference pattern in the topmost detector to vanish since we now \emph{can} identify whether the detected photon were created on the first or the second pass. The point here is that the interference pattern in the lower detector \emph{also}
vanishes in spite of the fact that we have done nothing to the paths relevant to this detector. How is this possible? The answer is entanglement. When the photons are created, they are created in entangled pairs. This means that they are correlated regardless of their spatial separation. Due to conservation of momentum when they are created the photons will deflect symmetrically and go to separate ``arms'' of the experimental setup. The deep insight of this is the one of non-locality, since we could, in principle make the arms arbitrary long and still get an \emph{instantaneous} effect in the lower arm due to the action at a distance. To push it further, we erase the knowledge we have regarding the paths of the photons by inserting a polarizer with an $45^o$ orientation in front of the topmost detector. This will obviously lead to the impossibility of distinguishing the photons, since a photon transmitted through the polarizer could equally well have taken either path. If we compare the obtained data from the two detectors \emph{a posteriori}, we will note an absolute correlation when we obtain and erase the which-path information. Now we see that interference is a manifestation of non-locality! One also notes that the consequence of entanglement is non-locality.

\section{The Bell inequality}
\subsection{Introduction}
Arguably, one of the most important conceptual discoveries in modern physics is the Bell inequality. It was derived by Bell, with the EPR \cite{EPR1,EPR2} thought experiment in mind to show that a \emph{local} hidden-variable theory cannot reproduce all the quantum mechanical results. The inequality is derived by making assumptions of a local reality. Empirical results, which are in excellent agreement with quantum mechanics, do violate the inequality thus refuting a local reality. Over the years the inequalities and the corresponding experiments have been progressively refined. Having said this, we should emphasize that there are some (as it should be, given the consequences) objections on whether the empirical evidence is conclusive \cite{Franson1}. However the loopholes for a local reality are successivly getting smaller and smaller.

\subsection{Empirical evidence of violation of Bell's inequality}
Since the late 1960's there have been propositions of inequalities which could be tested experimentally. It must be stressed that the example given above is not directly applicable to a real experiment, but has to be modified, however the main idea remains the same. The experiments has been consistent with quantum mechanics and heavily refuted the notion of local realism. Some ``loopholes'' for a local realism has been pointed out, however the experiments have become more and more refined thus ``narrowing'' the loopholes down considerably. An issue with detection has been pointed out but was addressed later on \cite{Rowe1}. Another objection called ``the light cone'' loophole has been made but again, has been addressed \cite{Weihs1}. An inequality designed to be compared to measurements is the Bell-CHSH inequality \cite{CHSH1}. We will make a derivation of the inequality for clarity.
Let us assume that A and B observes some random results from the sampling on the (hidden) variable $\lambda$. Furthermore, the results obtained by A and B are only dependent of the local detector setting and the common variable $\lambda$. We denote the value observed by A with setting a as A(a,$\lambda$), and analogously for B. Moreover, we consider a correlation of observables O and P as $C(O,P)=E(OP)$, where E is the expectation value
\begin{equation}
E(O)=\int_{\Lambda}{O(\lambda)\rho(\lambda)}
\end{equation}
,where $\rho(\lambda)$ is a probability density. Now, for clarity we assume that the only values obtained from a measurement is $\pm1$. However the derivation below is valid $\forall A,B\in [-1,1] $. Thus at least one of the two expressions
\begin{equation}
B(b,\lambda)+B(b',\lambda),\hspace{1cm}B(b,\lambda)-B(b',\lambda)
\end{equation}
has to be 0. This gives 
\begin{eqnarray}
&&A(a,\lambda)B(b,\lambda)+A(a,\lambda)B(b',\lambda)+A(a',\lambda)B(b,\lambda)-A(a',\lambda)B(b',\lambda)\nonumber\\
&\leq& |A(a,\lambda)(B(b,\lambda)+B(b',\lambda))+A(a',\lambda)(B(b,\lambda)-B(b',\lambda))|\nonumber\\
&\leq& |A(a,\lambda)(B(b,\lambda)+B(b',\lambda))|+|A(a',\lambda)(B(b,\lambda)-B(b',\lambda))|\nonumber\\
&\leq& |B(b,\lambda)+B(b',\lambda)|+|B(b,\lambda)-B(b',\lambda)|\leq 2.
\end{eqnarray}
Now, in terms of correlations, we get
\begin{equation}
F=|C(A(a),B(b))\pm C(A(a),B(b'))|+|C(A(a'),B(b))\mp C(A(a'),B(b'))|\leq2.
\label{CHSH}
\end{equation}
This is the CHSH inequality.\\ 
Now, let us consider the quantum mechanical predictions. The correlation of a pair of commuting $[A,B]=0$ observables is
\begin{equation}
\braket{AB}=\braket{AB\phi|\phi}.
\end{equation}
We consider an experiment where the spin of an electron is made. The settings a and a' corresponds to a measurement of the spin along the x or z-axis. The observable are represented by the Pauli matrices \cite{Sakurai1}
\begin{equation}
S_x=\left[\begin{array}{ccc}
0&1\\
1&0
\end{array}\right]
,\hspace{1cm}S_z=\left[\begin{array}{ccc}
1&0\\
0&-1
\end{array}\right].
\end{equation}
We denote the eigen-kets for $S_x$ as $\ket{\uparrow}, \ket{\downarrow}$. The situation in question is described by a spin singlet state
\begin{equation}
\ket{\phi}=\frac{1}{\sqrt{2}}(\ket{\uparrow}_1\ket{\downarrow}_2-\ket{\downarrow}_1\ket{\uparrow}_2).
\end{equation}
We let observer B rotate his system relative A by $45^o$ such that
\begin{eqnarray}
A(a)&=&S_z\nonumber\\
A(a')&=&S_x\nonumber\\
B(b)&=&-\frac{1}{\sqrt{2}}(S_z+S_x)\nonumber\\
B(b')&=&\frac{1}{\sqrt{2}}(S_z-S_x).
\end{eqnarray} 
Calculating the correlations, we get 
\begin{equation}
\braket{A(a)B(b)}+\braket{A(a)B(b')}+\braket{A(a')B(b)}-\braket{A(a')B(b')}=2\sqrt{2}.
\end{equation}

\begin{figure}[hbt!]
\begin{center}
\includegraphics[scale=0.4,angle=0]{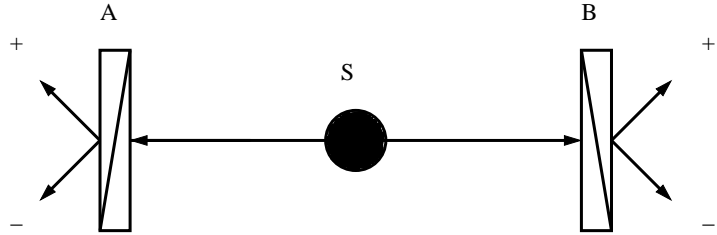} 	
\end{center}
\caption{Setup of the EPR \emph{gedanken experiment}. Two entangled spin-$\frac{1}{2}$ particles are described by a spin singlet state and emitted from a source S. They travel in opposite directions from the source and the spin components are measured at A and B. The coincidence rates are measured and compared \emph{a posteriori} and experimental results agree with predictions made by quantum mechanics, i.e. stronger correlations are experienced than would be predicted by a local theory.}		  
\end{figure}

This is often referred to as Tsirelson's bound. The operators giving this value are isomorphic to the Pauli matrices. 
The CHSH inequality is thus violated giving $F_{max}=2\sqrt{2}$. An early result is $F=2.697\pm 0.015$ \cite{Aspect1,Aspect2,Aspect3}. Among later results are $2.25\pm0.03$ \cite{Rowe1} and $2.92\pm 0.18$ for a strictly relativistic setup \cite{Weihs1, Tittel1}.

\subsection{Consequences of violation of Bell's inequalities}
Given the rather strong evidence, we accept that the Bell inequalities \emph{are} violated. The common interpretation of the violation of the Bell inequalities is that there is no possibility of a local reality for entangled objects. However, the non-locality cannot be used for superluminal communication, since from A's viewpoint, only random outcomes of the experiment is noted and A can never determine whether B has changed its experimental setting or not. It is only when collecting data \emph{a posteriori} that the (absolute) \emph{correlation} is noted, indicating the action at a distance \cite{Aspect4}. This argument is ``saving'' special relativity, or creates a ``peaceful coexistence between quantum mechanics and relativity'' (A.Shimony). The experimental results, and the theoretical implications of non-locality are truly mind boggling and raise truly fundamental questions of the nature of reality and physical connection. \\

\renewcommand\thechapter{Appendix B}
\renewcommand{\theequation}{B-\arabic{equation}}
\setcounter{equation}{0}
\renewcommand{\thesection}{B.\arabic{section}}

\appendix{}
\chapter{}
\section{The role of consciousness }

The notion of reality certainly has to be carefully established in the context of a collapse theory. While it may seem, at best, like an academic exercise in the classical realm, it becomes an absolutely fundamental one in quantum physics. Already when we made the postulation of the non-local interaction-induced reduction, we more or less made the tacit assumption of a subjective reality, i.e., giving in some sense an observer (or participator) driven reality. The contrary view seems to be in direct conflict with quantum mechanics and established empirical results. This however, presents us with new problems. Indeed it is a standpoint associated with great difficulties. Not least due to the self referring elements that seem unavoidable.\\ 
 Examining the consequences of our standpoint of a non-local, subjective, reality, we envision an universe which materializes upon observation or ``participation''.  So in quoting Heisenberg: ''The path of an electron \emph{only} exist when we observe it'', we adopt the view that there is no reality without observation.\\
Given that it is the linearity that is the ``culprit'' in a reduction context, it follows that the reduction process is inherently non-linear. Now, what is \emph{not} so clear is what the cause of the non-linearity is. Although the collapse process has been introduced in a standard interaction-type of way in this thesis, we do not consider it likely that it can be viewed as a standard quantum interaction, since linear quantum mechanics seem so successful in describing all kinds of interactions. The cause of the reduction-interaction has to be something fundamentally different from usual physical interactions.\\
   We believe that one has to extend the very notion of what is ``physical'' to even begin to contemplate the cause of the reduction process. There have been quite a few propositions of the involvement of consciousness in a reduction context. A (very dangerous) justification of this stance has been that there are some tenuous points of contacts between consciousness and quantum mechanics. Given the assumption that our consciousness belong to our reality, one can argue that the consciousness is in some way governed by (some sort of) ``quantum mechanics'', or more general; ``physical'' laws. The application of QM is certainly not so straightforward on consciousness! Objections are often raised, purely on emotional grounds that the ``act'' of consciousness is \emph{not} a physical one. This standpoint has to be considered a bit naive, since, really, consciousness eventually \emph{is} the foundation of science and certainly a reality! (As all our experimental and theoretical results ultimately are filtered through our consciousness.) \\
Now, we observe the law of action and reaction in nature at all levels. Is consciousness different in that respect? It can certainly be acted upon (i.e. awareness), so in that perspective it would be \emph{unlikely} that the consciousness \emph{cannot} act, but only be acted upon. Given this reasoning, a  hypothesis could be that reality is fundamentally subjective, i.e., it is only through our conscious observation, or participation the nature realizes itself. However the action of consciousness \emph{is} fundamentally \emph{different} from any other physical action. In the language of this thesis, we call this difference (quantum) non-linearity. So the collapse-division is, under this hypothesis, not between ``small'' and ``big'', but between mind and matter. We see this as a very intriguing possibility, since, we are following the causation-chain from the electron to the screen, the retina of the observer to the brain with its neurons, without ever encountering something fundamentally different which could be responsible for the non-linearity. But persisting in following this chain, we finally arrive at, what we here choose to call consciousness. The very point is that, since the collapse is fundamentally nonlinear it also has to originate from something entirely different than usual interactions, which we (apparently) can describe perfectly well with linear interactions.
Some first attempts have been made of testing whether or not consciousness \emph{is} responsible for reducing the quantum mechanical state vector \cite{Bierman1}. Although the experiment mentioned does agree with a subjective conscious reduction it is certainly far from conclusive. Given today's technology it should be possible to make further refinements of the experiment. And moreover explain the discrepancies between experiments.\\ 
 The task of examining this possibility is truly daunting. Still, or perhaps just because of the complexity, we find further studies of the possibility of a mind-matter interaction worthwhile. So we end this section by quoting J.S. Bell: ``As regards mind, I am fully convinced that it has a central place in the ultimate nature of reality'' \cite{Bell1}.



\end{document}